\documentclass[a4paper]{article}
\usepackage{iwslt17,amssymb,amsmath,epsfig}
\usepackage{graphicx,tikz,booktabs,float}
\usetikzlibrary{arrows}
\usepackage[utf8]{inputenc}
\usepackage{pgfplots}
\pgfplotsset{width=10cm,compat=1.9}
\def\reg{{\rm\ooalign{\hfil
     \raise.07ex\hbox{\scriptsize R}\hfil\crcr\mathhexbox20D}}}

\title{Phonemic and Graphemic Multilingual CTC Based Speech Recognition}

\name{Markus Müller$^1$, Sebastian Stüker$^1$, Alex Waibel$^{1,2}$}
\address{$^1$Interactive Systems Lab, Institute for Anthropomatics and Robotics  \\
Karlsruhe Institute of Technology, Karlsruhe, Germany \\
$^2$Carnegie Mellon University, Pittsburgh PA, USA \\
{\small \tt m.mueller@kit.edu}
}
\begin{document}
\maketitle
\begin{abstract} % 173 words
Training automatic speech recognition (ASR) systems requires large amounts of data in the target language in order to achieve good performance.
Whereas large training corpora are readily available for languages like English, there exists a long tail of languages which do suffer from a lack of resources.
One method to handle data sparsity is to use data from additional source languages and build a multilingual system.
Recently, ASR systems based on recurrent neural networks (RNNs) trained with connectionist temporal classification (CTC) have gained substantial research interest.
In this work, we extended our previous approach towards training CTC-based systems multilingually.
Our systems feature a global phone set, based on the joint phone sets of each source language.
We evaluated the use of different language combinations as well as the addition of Language Feature Vectors (LFVs).
As contrastive experiment, we built systems based on graphemes as well.
Systems having a multilingual phone set are known to suffer in performance compared to their monolingual counterparts.
With our proposed approach, we could reduce the gap between these mono- and multilingual setups, using either graphemes or phonemes.
\end{abstract}

% \begin{keywords}
%One, two, three, four, five
% speech recognition, multilingual, CTC
% \end{keywords}

\section{Introduction}
\label{sec:intro}
Automatic speech recognition systems have matured dramatically in recent years, lately with reported recognition accuracies similar to those of humans on certain tasks \cite{xiong2016achieving,xiong2017microsoft}.
A large amount of carefully prepared training data is required to achieve this level of performance.
While such data is available for well-researched and -resourced languages like English, there exists a long tail of languages for which such training material does not exist.
Various methods have been proposed to handle data sparsity.
In this work, we focus on multilingual systems:
A common approach is to incorporate data from supplementary source languages in addition to data from the target language.

Lately, systems based on RNNs trained with connectionist temporal classification (CTC) \cite{graves2006connectionist} have become popular.
In this work we focus on building multilingual RNN/CTC systems, instead of systems based on either GMM/HMM or DNN/HMM, with the goal of applying them in a multi-lingual manner and are planning crosslingual experiments in the future. 
For this future crosslingual case, the multilingual RNN acts as a network that can be adapted to multiple languages for which only very little adaptation data is available.
In the multilingual scenario of this paper, we have one multilingual model that is able to recognize speech from multiple languages simultaneously, while for all languages a comparatively large amount of training data is available.
This is particular useful in environments with fast language changes.

Recently, we demonstrated the use of a second language in addition to the target language when building a phoneme based CTC system \cite{mueller2017specom}.
We now extend this approach by using data from up to 4 languages (English, French, German and Turkish).
% We used a joint multilingual model with a universal phone set.
Building systems using phones as acoustic modeling unit requires a pronunciation dictionary.
But, creating these dictionaries is a time-consuming, resource intense process and often a bottle-neck when building speech recognition systems for new languages.
While automatic methods to create pronunciations for new words given an existing dictionary exist \cite{bisani2008joint,novak2011phonetisaurus}, such approaches are based on an existing seed dictionary.
Using graphemes as acoustic modeling units, instead has the advantage of loosing the need for a pronunciation dictionary at the cost that graphemes might not always be a good modeling unit, depending on the grapheme-to-phoneme relation of the target language. \cite{ip_schillo2000, ip_kanthak2002, ip_killer2003}
This is particularly challenging in a multilingual setting, because different languages, although they might share the same writing system, do feature different pronunciation rules \cite{ip_kanthak2003, ip_stueker2008a, ip_stueker2008b}.

This paper is organized as follows: Next, in Section (\ref{sec:rework}), we provide an overview of related work in the field.
In Section \ref{sec:mlctc}, we describe our proposed approach, followed by the experimental setup in Section \ref{sec:expsetup}.
The results are presented in Section \ref{sec:results}.
This paper concludes with Section \ref{sec:conclusion}, where we also provide an outlook to future work.
\section{Related Work}
\label{sec:rework}
\subsection{Multi- and Crosslingual Speech Recognition Systems}
\label{sec:relwork:subsec:gmm}
Using GMM/HMM based systems was considered state of the art prior to the emergence of systems with neural networks.
Data sparsity has been addressed in the past, by training systems multi- and crosslingually \cite{wheatley1994evaluation,schultz1997fast}.
Methods for crosslingual adaptation exist \cite{stuker2009acoustic}, but also methods for adapting the cluster tree were proposed \cite{schultz2001language}.
% The use of ML-mix, ML-seg or ML-tag explicitly models how data is being shared between languages, whereas our proposed RNN-based approach does this implicitly.
Traditional systems typically use context-dependent phones.
When trained multi- or crosslingually, the clustering of phones into context-dependent phones needs to be adapted \cite{schultz2000polyphone}. %\cite{vu2014multilingual}.

But when using an RNN, the system is trained on context-independent targets, so that in the multilingual case this kind of adaptation is unnecessary, as the network learns the context-dependency during training.
%This does not require to adapt the set of context-dependent phones as the temporal dependencies are being learned implicitly by the recurrent network.
%
\subsection{Multilingual Bottleneck Features}
Deep Neural Networks (DNNs) are a data-driven method with many parameters to be trained, failing to generalize if trained on only a limited data set.
Different methods have been proposed to train networks on data from multiple source languages.
Training DNNs typically involves a pre-training and a fine-tuning step.
It has been shown, that the pre-training is language independent \cite{SwietojanskiGR12}.
Several approaches exist to fine-tune a network using data from multiple languages.
One method is to share hidden layers between languages, but to use language specific output layers \cite{ghoshalmultilingual,scanzio,ip_heigold2013,VeselyKGJE12}.
Combining language specific output layers into one layer is also possible \cite{grezl2014adaptation}.
By dividing the output layer into language specific blocks, the setup uses language dependent phone sets.
Training DNNs simultaneously on data from multiple languages on the other hand can then be considered a form of multi-task learning \cite{caruana1997multitask,mohan2015multi}.
\subsection{Neural Network Adaptation}
By supplying additional input features, neural networks can be adapted to various conditions.
One of the most common methods is to adapt neural nets to different speakers by providing a low dimensional code representing speaker characteristics.
These so called i-Vectors \cite{saon2013speaker} allow to train speaker adaptive neural networks \cite{miao2014towards}.
An alternative method for adaptation are Bottleneck Speaker Vectors (BSVs) \cite{huang2015investigation}.

Similar to BSVs, we proposed an adaptation method for adapting neural networks to different languages when trained on multiple languages.
We first proposed using the language identity information via one-hot encoding \cite{mueller2015}.
One of the shortcomings of this approach is that it does not supply language characteristics to the network.
To address this issue, we proposed Language Feature Vectors (LFVs) \cite{mueller2016,itgmueller2016} which have shown to encode language properties, even if the LFV net was not trained on the target language.
\subsection{CTC Based ASR Systems}
Recently, RNN-based systems trained using the CTC loss function \cite{graves2006connectionist} have become popular.
Similar to traditional ASR systems, CTC based ones are trained using either phones, graphemes, or both \cite{chen2014joint}.
Training on units larger than characters is also possible \cite{sennrich2015neural}.
This method, called Byte Pair Encoding (BPE), derives larger units based on the transcripts.
Given enough training data, even training on whole words is possible \cite{soltau2016neural}.
Multi-task learning has also been proposed \cite{kim2016joint,lu2017multi,sak2017multi}.
CTC based systems are able to outperform HMM based setups on certain tasks \cite{miao2016empirical}.
%We proposed a first approach towards training CTC systems multilingually \cite{mueller2017specom}.
%In this work, we use a network architecture based on Baidu's Deepspeech2 setup \cite{amodei2016deep}.
%
\section{Language Adaptive Multilingual CTC Based Systems}
\label{sec:mlctc}
Traditional speech recognition systems typically rely on a pronunciation dictionary which maps words to phone sequences.
It is also possible to train systems on graphemes as acoustic units, but this affects the performance depending on the language.
While there are languages with a close mapping between letters and sounds, e.g., Spanish, this does not hold for every language.
Pronunciation rules are quite complex, with groups of characters being mapped to different sounds based on their context.
An example of such complex mappings would be English. The string ``ough'' has 8 different acoustic realizations, depending on the context, as in, e.g., ``rough'', ``ought'' or ``through''.
\subsection{Multilingual Systems}
Speech recognition systems are typically built to recognize speech of a single language.
Training traditional systems multilingually involves a hybrid DNN/HMM setup where the hidden layers of the DNN are shared between languages and the output layers are kept language dependent.
Such systems can be seen as individual, language dependent systems, trained jointly. % together.
Training language universal systems using a global phones set is possible, however HMM based systems do not generalize well when being trained on multiple languages.
In this work, we propose an approach using RNN based systems trained using CTC on data from multiple languages, with a global set of units modeling the acoustics (graphemes or phones).
The main advantage of such a system is the ability to recognize speech from multiple languages simultaneously, without knowledge of the input language's identity.

In the past, we proposed a setup for training CTC-based systems multilingually using a universal phone set \cite{mueller2017specom}.
In this work, we extended our previous work in three ways:
1) we increased the number of languages used
2) we used multilingually trained bottleneck features (BNFs)
3) in addition to phones, we evaluated the use of graphemes.
In the past, we demonstrated the use of LFVs using DNN/HMM-based systems for multilingual speech recognition.
We now apply this technique to CTC-based speech recognition.
\subsection{Language Feature Vectors}
\label{subsec:lfv_theory}
LFVs are a low dimensional representation of language properties, extracted using a neural network.
The setup consisted of two networks, Figure \ref{fig:lidnn} shows the network architecture.
The first network was used to extract BNFs from acoustic input features.
It was trained using a combination of lMel and tonal features as input and phone states as targets.
The second network was trained for language identification using BNFs as input features.
In contrast to networks trained for speech recognition, we used a much larger input context because of the language information being long-term in nature.
This network was trained to detect languages and featured a bottleneck layer, which was used to extract the LFVs after training.
\tikzstyle{layer}=[draw=black,fill=black!30]
\tikzstyle{layerlid}=[draw=black,fill=green!30]
\tikzstyle{dots}=[draw=black,fill=black]
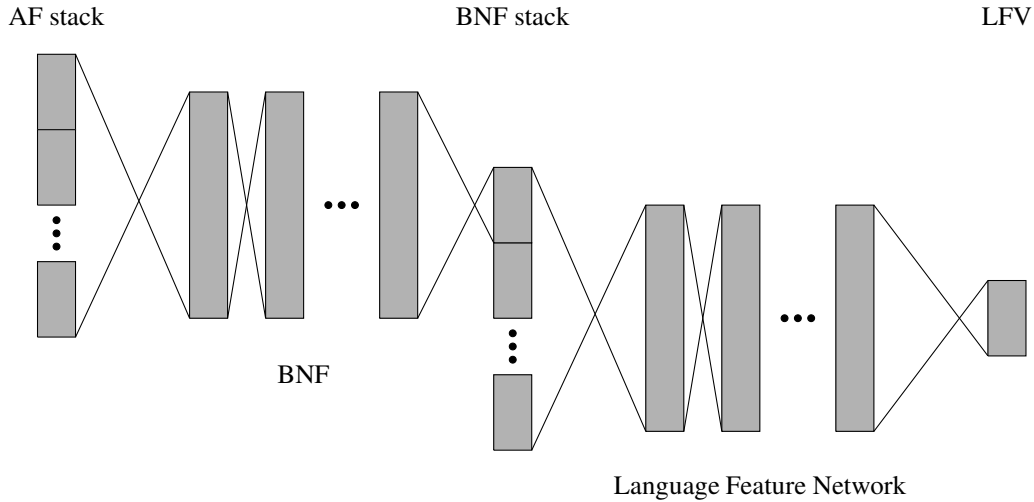
\begin{figure*}[htbp]
  \centering

  \begin{tikzpicture}[scale=1.0]

  % Acoustic feature stack

  \draw (-1.75, 2.5) node[draw=white,fill=white] {AF stack};

  \fill[layer] (-2,1) coordinate(l0bl) -- (-1.5,1) coordinate(l0br) -- (-1.5,2) coordinate(l0tr) -- (-2,2) coordinate(l0tl) -- (-2,1);

  \fill[layer] (-2,0) coordinate(l0_1bl) -- (-1.5,0) coordinate(l0_1br) -- (-1.5,1) coordinate(l0_1tr) -- (-2,1) coordinate(l0_1tl) -- (-2,0);

  \draw[dots] (-1.75,-0.2) circle (0.045);
  \draw[dots] (-1.75,-0.375) circle (0.045);
  \draw[dots] (-1.75,-0.55) circle (0.045);

  \fill[layer] (-2,-1.75) coordinate(l0_2bl) -- (-1.5,-1.75) coordinate(l0_2br) -- (-1.5,-0.75) coordinate(l0_2tr) -- (-2,-0.75) coordinate(l0_2tl) -- (-2,-1.75);

  % Bottle-neck stack

  \fill[layer] (4,-1.5) coordinate(l5_1bl) -- (4.5,-1.5) coordinate(l5_1br) -- (4.5,-0.5) coordinate(l5_1tr) -- (4,-0.5) coordinate(l5_1tl) -- (4,-1.5);

  \draw[dots] (4.25,-1.7) circle (0.045);
  \draw[dots] (4.25,-1.875) circle (0.045);
  \draw[dots] (4.25,-2.05) circle (0.045);

  \fill[layer] (4,-3.25) coordinate(l5_2bl) -- (4.5,-3.25) coordinate(l5_2br) -- (4.5,-2.25) coordinate(l5_2tr) -- (4,-2.25) coordinate(l5_2tl) -- (4,-3.25);

  \draw (1.5,-2.25) node {BNF};

  \draw (4.25,2.5) node[draw=white,fill=white] {BNF stack};

  % BNF Layers

  \fill[layer] (4,-0.5) coordinate(l5bl) -- (4.5,-0.5) coordinate(l5br) -- (4.5,0.5) coordinate(l5tr) -- (4,0.5) coordinate(l5tl) -- (4,-0.5);
  \fill[layer] (2.5,-1.5) coordinate(l4bl) -- (3,-1.5) coordinate(l4br) -- (3,1.5) coordinate(l4tr) -- (2.5,1.5) coordinate(l4tl) -- (2.5,-1.5);

  \draw[dots] (2,0) circle (0.045);
  \draw[dots] (2.175,0) circle (0.045);
  \draw[dots] (1.825,0) circle (0.045);

  \fill[layer] (1,-1.5) coordinate(l2bl) -- (1.5,-1.5) coordinate(l2br) -- (1.5,1.5) coordinate(l2tr) -- (1,1.5) coordinate(l2tl) -- (1,-1.5);
  \fill[layer] (0,-1.5) coordinate(l1bl) -- (0.5,-1.5) coordinate(l1br) -- (0.5,1.5) coordinate(l1tr) -- (0,1.5) coordinate(l1tl) -- (0,-1.5);

  \draw (l0tr) -- (l1bl);
  \draw (l0_2br) -- (l1tl);
  \draw (l1tr) -- (l2bl);
  \draw (l1br) -- (l2tl);
  \draw (l4tr) -- (l5bl);
  \draw (l4br) -- (l5tl);

  % DNN Layers

  \fill[layer] (6,-3) coordinate(l6bl) -- (6.5,-3) coordinate(l6br) -- (6.5,0) coordinate(l6tr) -- (6,0) coordinate(l6tl) -- (6,-3);
  \fill[layer] (7,-3) coordinate(l7bl) -- (7.5,-3) coordinate(l7br) -- (7.5,0) coordinate(l7tr) -- (7,0) coordinate(l7tl) -- (7,-3);

  \draw[dots] (8,-1.5) circle (0.045);
  \draw[dots] (8.175,-1.5) circle (0.045);
  \draw[dots] (7.825,-1.5) circle (0.045);

  \fill[layer] (8.5,-3) coordinate(l11bl) -- (9,-3) coordinate(l11br) -- (9,0) coordinate(l11tr) -- (8.5,0) coordinate(l11tl) -- (8.5,-3);
  \fill[layer] (10.5,-2) coordinate(l12bl) -- (11,-2) coordinate(l12br) -- (11,-1) coordinate(l12tr) -- (10.5,-1) coordinate(l12tl) -- (10.5,-1);

  \draw (10.75,2.5) node[draw=white,fill=white] {LFV};

  \draw (l5tr) -- (l6bl);
  \draw (l5_2br) -- (l6tl);

  \draw (l6tr) -- (l7bl);
  \draw (l6br) -- (l7tl);
  \draw (l11tr) -- (l12bl);
  \draw (l11br) -- (l12tl);

  \draw (l12tl) -- (l12bl);

  \draw (7.5,-3.75) node {Language Feature Network};

  \end{tikzpicture}
  \caption{Overview of the network architecture used to extract language feature vectors (LFV). The acoustic features (AF) are being pre-processed in a DBNF in order to extract BNFs. These BNFs are being stacked and fed into the second network to extract LFVs.}
  \label{fig:lidnn}
\end{figure*}
\subsection{Input Features}
Using BNFs as input features is common for traditional speech recognition systems.
By forcing the information to pass through a bottleneck, the network creates a low-dimensional representation of features relevant to discriminate between phones.
DNN/HMM or GMM/HMM based systems benefit from using such features over plain features like, e.g., MFCCs.
We evaluated training our CTC systems on multilingual BNFs.
\subsection{Network Architecture}
The network architecture chosen was based on Baidu's Deepspeech2 \cite{amodei2016deep}.
As shown in Figure \ref{fig:ctcnn}, the network consists of two TDNN / CNN layers.
We add LFVs to the output of the second TDNN / CNN layer as input to the bi-directional LSTM layers.
We use a feed-forward output layer to map the output of the last LSTM layer to the targets.
\tikzstyle{layer}=[draw=black,fill=black!30]
\tikzstyle{layerlfv}=[draw=black,fill=green!30]
\tikzstyle{layerrnn}=[draw=black,fill=blue!30]
\tikzstyle{layercnn}=[draw=black,fill=orange!30]
\tikzstyle{arrow} = [semithick,fill=red!30,line width=1.4pt, shorten >= 4.5pt]
\tikzstyle{dots}=[draw=black,fill=black]
%\begin{figure}[htbp]
\begin{figure}[!h]
\centering
\begin{tikzpicture}[scale=0.5]

% CNN layer
\fill[layercnn] (-6,3) coordinate(c1tl) -- (-7,3) coordinate(c1tr) -- (-7,-3) coordinate(c1br) -- (-6,-3) coordinate(c1bl) -- (-6,3);

\fill[layercnn] (-4,2.5) coordinate(c2tl) -- (-5,2.5) coordinate(c2tr) -- (-5,-2.5) coordinate(c2br) -- (-4,-2.5) coordinate(c2bl) -- (-4,2.5);

% LFV layer
\fill[layerlfv] (-4,3.5) coordinate(lfvtl) -- (-5,3.5) coordinate(lfvtr) -- (-5,2.5) coordinate(lfvbr) -- (-4,2.5) coordinate(lfvbl) -- (-4,3.5);

% LSTM layer
\fill[layerrnn] (-2.5,4) coordinate(l1tl) -- (-1.5,4) coordinate(l1tr) -- (-1.5,-4) coordinate(l1br) -- (-2.5,-4) coordinate(l1bl) -- (-2.5,4);

\fill[layerrnn] (-0.5,4) coordinate(l2tl) -- (0.5,4) coordinate(l2tr) -- (0.5,-4) coordinate(l2br) -- (-0.5,-4) coordinate(l2bl) -- (-0.5,4);

\fill[layerrnn] (1.5,4) coordinate(l3tl) -- (2.5,4) coordinate(l3tr) -- (2.5,-4) coordinate(l3br) -- (1.5,-4) coordinate(l3bl) -- (1.5,4);

\fill[layerrnn] (3.5,4) coordinate(l4tl) -- (4.5,4) coordinate(l4tr) -- (4.5,-4) coordinate(l4br) -- (3.5,-4) coordinate(l4bl) -- (3.5,4);

% Output layer
\fill[layer] (6,2) coordinate(outtl) -- (7,2) coordinate(outtr) -- (7,-2) coordinate(outbr) -- (6,-2) coordinate(outbl) -- (6,2);

% Descriptions
\node[align=center,font=\small,rotate=0] at (-5.5,-5) {2D Convolution\\Layers};
\node[align=center,font=\small,rotate=0] at (1,-5) {Bi-directional LSTM Layers};
\node[align=center,font=\small,rotate=0] at (6.5,-5) {Output\\Layer};
\node[align=center,font=\small,rotate=0] at (-4.5,4.5) {LFV};

% Links between networks

% c1 - c2
\draw (c1bl) -- (c2br);
\draw (c1tl) -- (c2tr);

% c2 - l1
\draw (c2bl) -- (l1tl);
%\draw (c2tl) -- (l1bl);

% lfv - l1
%\draw (lfvbl) -- (l1tl);
\draw (lfvtl) -- (l1bl);

% l1 - l2
\draw (l1br) -- (l2tl);
\draw (l1tr) -- (l2bl);

% l2 - l3
\draw (l2br) -- (l3tl);
\draw (l2tr) -- (l3bl);

% l3 - l4
\draw (l3br) -- (l4tl);
\draw (l3tr) -- (l4bl);

% l4 - out
\draw (l4br) -- (outtl);
\draw (l4tr) -- (outbl);

\end{tikzpicture}
\caption{Network layout, based on Baidu's Deepspeech2 \cite{amodei2016deep}. LFVs are being added after the final convolution layer.}
\label{fig:ctcnn}
\end{figure}
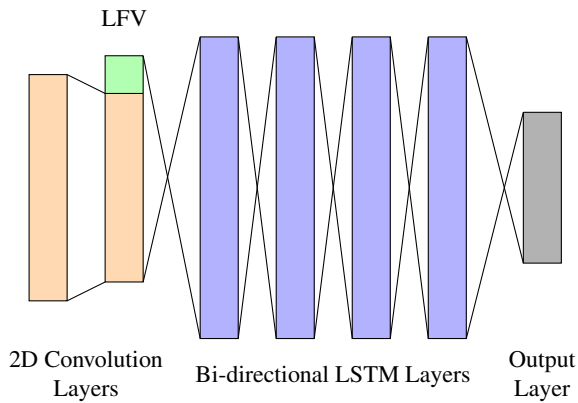
\section{Experimental Setup}
\label{sec:expsetup}
We built our systems using a framework based on PyTorch \cite{pytorch},
%which provided Python bindings to Torch \cite{torch}, 
as well as warp-ctc \cite{warpctc} for computing the CTC loss during network training.
To extract acoustic features from the data, we used the Janus Recognition Toolkit (JRTk) \cite{woszczyna1993}, which features the IBIS single-pass decoder \cite{soltau2001one}.
\subsection{Dataset}
% Part 1: Corpus
We conducted our experiments using data from the Euronews Corpus \cite{gretter2014euronews}, a dataset containing recordings of TV broadcast news from 10 different languages (Arabic, English, French, German, Italian, Polish, Portuguese, Russian, Spanish, Turkish), with orthographic transcripts at utterance level.
The advantage of this dataset is that the channel conditions do not differ between languages, ensuring that we are adapting our systems to different languages instead of different channel conditions, like, e.g., different environmental noises present in different languages.
We filtered the available data, % based on the length of utterances.
retaining only utterances with a length of at least 1s and a transcript length of at most 639 symbols, because of an internal limitation within CUDA\footnote{see: https://github.com/baidu-research/warp-ctc, accessed 2017-10-09}.

Noises were annotated in a very basic way, consisting of only one generic noise marker covering both human and non-human noises.
With noises accounting for a quite large amount of utterances, we only selected a small subset of them to account for a more balanced set of training data.
After applying all filtering steps, approximately 50h of data per language was available.
We split the available data on a speaker basis into a 45h training and 5h test set.
\subsection{Acoustic Units}
We conducted experiments using both phones and graphemes as acoustic units.
As graphemes we used the provided transcripts, while we used MaryTTS \cite{schroder2003german} to generate a pronunciation dictionary automatically to map words to phones.
In addition, we included a marker to indicate word boundaries.
\subsection{Input Features}
\label{subsec:inputfeatures}
As input features, we used log Mel and tonal features (FFV \cite{kornel:ffv} and pitch \cite{kjell:da}), extracted using a 32ms window with a 10ms frame-shift.
We included tonal features as part of our standard pre-processing pipeline because previous experiments showed a reduction in the word error rate (WER) of speech recognition systems, even if the language is not tonal \cite{metze2013models}.

Based on these features, we trained a network for extracting multilingual bottleneck features (BNFs).
The network featured 5 feed-forward layers, with 1,000 neurons per layer, with the second last layer being the bottleneck with only 42 neurons.
The acoustic features were fed with a context of $+/-$ 6 frames into the network.
While the hidden layers were shared between languages, we used language dependent output layers.
6,000 context-dependent phone states were used as targets, with data from 5 languages (French, German, Italian, Russian, Turkish).
To obtain phone state labels, DNN/HMM systems for each language were trained.
After training, all layers after the bottleneck were discarded and the output activations of this layer were taken as BNFs.
\subsection{LFV Network Training}
Training the network for the extraction of LFVs is a two step process.
First, BNFs are being trained (see Section \ref{subsec:inputfeatures}), and then based on these BNFs, a second network is trained to recognize the language.
This network features 6 layers with 1,600 neurons per layer, except for the bottleneck layer with only 42 neurons.
In contrast to networks trained for speech recognition, this network featured a large context spanning $+/-$ 33 frames.
To reduce the dimensionality of the input, only every third frame was taken.
For training this network, we used data from 9 languages( all available languages in the corpus except English).
%We omitted English as we wanted to asses the performance of our system in a crosslingual manner on a language not seen during training.
%
\subsection{CTC RNN Network Training}
The RNN network was trained using either log Mel / tonal features or BNFs.
As targets, we used both graphemes and phonemes as acoustic units, with an additional symbol added for separating words.
% The network architecture chosen was based on Baidu's Deepspeech2 \cite{amodei2016deep}.
% As shown in Figure \ref{fig:ctcnn}, the network consists of two TDNN / CNN layers.
The networks were trained using stochastic gradient descent (SGD) with Nesterov momentum \cite{sutskever2013importance} of 0.9 and a learning rate of 0.0003.
Mini-batch updates with a batch size of 20 and batch normalization were used.
%To prevent gradients from exploding, a max norm constraint of 400 was enforced.
During the first epoch, the network was trained with utterances sorted ascending by length to stabilize the training, as shorter utterances are easier to align.
\subsection{Evaluation}
To evaluate our setup, we used the same decoding procedure as in \cite{graves2006connectionist} and greedily search the best path without an external language model and evaluated our systems by computing the token error rate (TER) as primary measure.
In addition, we trained a character based neural network language model for English on the training utterances, as described in \cite{zenkel2017comparison}, so that for the recognition of English we could also measure a word error rate (WER) by decoding the network outputs with this language model. 
As the language model is only trained on only a small amount of data, the word error rate obtained with it should indicate  whether the improvements in TER of the pure CTC model measured on English also lead to a better word level speech recognition system.
\section{Results}
\label{sec:results}
%
% For our initial experiments, we used the same configuration as in \cite{mueller2017specom} by using only data from English and German.
% Based on those results, we then added data from more languages, using a combination of 4 languages (English, French, German, Turkish) in total.
We first evaluated using multilingual BNFs over plain log Mel / tone features.
Next, we used multilingual BNFs to train systems using a combination of 4 languages (English, French, German, Turkish).
\subsection{Multilingual BNFs}
First, we evaluated the use of multilingually trained BNFs as input features. % for our CTC based system.
To assess the performance, we trained systems for English and German monolingually on all available data.
The results are shown in Table \ref{tab:lmelbnf}.
The gain by the addition of BNFs is larger for German which can be explained by German being among the languages the BNF net was trained on (see Section \ref{subsec:inputfeatures}).
But the BNFs also show an improvement for English, although they did not see this language during training.
\begin{table}[h!]
\centering
\begin{tabular}{l|c|c}
\toprule
\textbf{Condition} & \textbf{English TER} & \textbf{German TER}  \\
\midrule
log Mel + Tone   & 13.0\% & 10.8\% \\
ML BNF        & 10.2\% &  7.8\% \\
\bottomrule
\end{tabular}
\caption{Comparison of using ML-BNFs over log Mel + tone features}
\label{tab:lmelbnf}
\end{table}
%
% \subsection{Low-resource}
% %
% Based on previous results, we evaluated the performance of our setup using only a subset of data per language.
% We limited ourselves to 8h per language, compared to 40h when using the full dataset.
% %The LER rises when limiting to only 8h of data.
% As shown in Table \ref{tab:lowresource}, the error rate rises by using less training data.
% %
% % DEBUG: length ids=50000 length=27.5h
% % Length of _tr_: 5.5h
% % Length of _en_: 7.4h
% % Length of _fr_: 7.8h
% % Length of _de_: 6.8h
% %
% \begin{table}[h!]
% \centering
% \caption{Monolingual comparison of using either 40h or 8h training data}
% \begin{tabular}{l|c|c}
% \toprule
% \textbf{Condition} & \textbf{English LER} & \textbf{German LER}  \\
% \midrule
% 40h   & 13.0\% & 10.8\% \\
% 8 h   & 21.0\% & 16.2\% \\
% \bottomrule
% \end{tabular}
% \label{tab:lowresource}
% \end{table}
%
\subsection{Multilingual Phoneme Based Systems}
Next, we evaluated the performance using 4 languages (English, French, German, Turkish). %, 8h per language, 32h of data in total.
We evaluated adding the LFVs after the TDNN / CNN layers. % or modulating the output of the second LSTM layer.
As baseline, we did not apply our language adaptation technique and used only multilingual BNFs.
%As shown in Table \ref{tab:lowphoneme}, modulating the output of the second LSTM layer results in the best multilingual performance.
As shown in Table \ref{tab:phoneme}, adding LFVs after the TDNN / CNN layer shows improvements over the baseline. %, but does not improve the performance as much as the modulation.
The relative improvements vary and while the language adapted systems are not en par with the monolingual ones, the adaptation does decrease the gap between the multi- and monolingual setup.
\begin{table}[h!]
\centering
\begin{tabular}{l|c|c|c|c}
\toprule
\textbf{Condition} & \textbf{DE} & \textbf{EN} & \textbf{FR} & \textbf{TR} \\
\midrule
Monolingual   & 7.8\% & 10.2\% & 8.3\% & 7.1\% \\
\midrule
ML  & 9.9\% & 14.1\% & 12.8\% & 8.4\% \\
ML + LFV & 8.9\% & 12.9\% & 10.7\% & 7.6\% \\
\bottomrule
\end{tabular}
\caption{Term Error Rate (\textbf{TER}) of multilingual (\textbf{ML}) phoneme CTC based systems, trained on 4 languages.}
\label{tab:phoneme}
\end{table}
%
% NOTE: Maybe re-add the next section
% To compare to our results to the initial setup, we trained a system using only data from German and English, 16h in total.
% As shown in Table \ref{tab:lowdeen}, improvements by using LFVs can be achieved in the same way as training on 4 languages, although the performance of the multilingual systems is higher compared to training on more languages.
% More languages introduce more ambiguity which makes it harder for the network to generalize.
% % This can be explained by more languages introducing a larger amount of ambiguity.
% The performance for German almost matched the monolingual one in this setup.
% %
% \begin{table}[h!]
% \centering
% \caption{Multilingual (\textbf{ML}) Phoneme CTC based systems, trained on 2 languages.}
% \begin{tabular}{l|c|c}
% \toprule
% \textbf{Condition} & \textbf{German (LER)} & \textbf{English (LER)}  \\
% \midrule
% Monolingual   & 16.2 & 21.0 \\
% \midrule
% ML Baseline   & 19.8 & 24.6 \\
% ML + LFV add  & 18.3 & 23.2 \\
% % ML + LFV mod & 16.5 & 22.6 \\
% \bottomrule
% \end{tabular}
% \label{tab:lowdeen}
% \end{table}
%
\subsection{Multilingual Grapheme Based Systems}
In addition to using phones, we also evaluated the performance using only the transcripts, without a pronunciation dictionary.
%The error rates are higher compared to phone based setups, but this expected as the units used for modelling the acoustics to not reflect the actual sounds.
As shown in Table \ref{tab:grapheme}, using LFVs improves the performance in this condition as well.
For English and French, the TER is higher compared to their phoneme counterpart, whereas lower TERs could be observed for both German and Turkish.
%The RNNs did perform better in modeling the pronunciations for Turkish and German compared to trained on phonemes directly.
% Turkish is a language known to have a relationship between letters and sounds that is close to 1:1.
%Turkish is a language having simple letter to sound rules.
%German, although having a more difficult mapping also does performed better given the transcripts alone.
%English, having complex pronunciation rules, does perform worse compared to the phonemic systems.
%The same phenomena could be observed for French.
One explanation could be that English and French feature more complex pronunciation rules that are better reflected by MaryTTS' language definitions.
The generated pronunciations for German and Turkish appear to worsen the performance.
The RNN seems to capture the letter to sound rules for these languages better.
%Turkish has letter to sound rules that are simpler compared to English and French.
%
\begin{table}[h!]
\centering
\begin{tabular}{l|c|c|c|c}
\toprule
\textbf{Condition} & \textbf{DE} & \textbf{EN} & \textbf{FR} & \textbf{TR} \\
\midrule
Monolingual   & 7.5\% & 12.9\% & 11.5\% & 6.6\% \\
\midrule
ML  & 9.1\% & 15.6\% & 13.4\% & 7.9\% \\
ML + LFV & 7.9\% & 14.3\% & 12.5\% & 7.3\% \\
\bottomrule
\end{tabular}
\caption{Term Error Rate (TER) of multilingual (\textbf{ML}) grapheme CTC based systems, trained on 4 languages.}
\label{tab:grapheme}
\end{table}

For English, we also trained a basic character based language model to decode the network output and compute the WER. As shown in Table \ref{tab:graphemewer}, similar improvements can be observed by adding LFVs. % to the network input.
\begin{table}[h!]
\centering

\begin{tabular}{l|c|c|c}
\toprule
\textbf{Condition} & \textbf{Mono} & \textbf{ML} & \textbf{ML + LFV} \\
\midrule
English   & 25.2\% & 30.8\% & 28.1\% \\
\bottomrule
\end{tabular}
\caption{Word Error Rate (\textbf{WER}) of English phoneme CTC based systems. Adding LFVs improves the multilingual performance.}
\label{tab:graphemewer}
\end{table}
\section{Conclusion}
\label{sec:conclusion}
We have presented an approach to adapt recurrent neural networks to multiple languages.
Using multilingual BNFs improved the performance, as well as providing LFVs for language adaptation.
These language adaptive networks are able to capture language specific peculiarities in a multilingual setup which results in an increased performance.
Such multilingual systems are able to recognize speech from multiple languages simultaneously.

Future work includes the use of different language combinations and working towards cross-lingual knowledge transfer.
We aim at further closing the gap between mono- and multilingual systems using additional adaptation techniques.
%
%%%%%%%%%%%%%%%%%%%%%%%%%%%%%%%%%%%%%%%%%%%%%%%%%%%%%%%%%%%%%%%%%%%%%%%
% End of paper
%%%%%%%%%%%%%%%%%%%%%%%%%%%%%%%%%%%%%%%%%%%%%%%%%%%%%%%%%%%%%%%%%%%%%%%
%
\bibliographystyle{IEEEtran}
\bibliography{refs}
\end{document}